\documentclass[prl,twocolumn,aps,superscriptaddress,showpacs,floatfix]{revtex4}
\usepackage{amsmath}
\usepackage{graphicx}
\begin{document}
\title{Demon-free quantum Brownian motors}
\author{L. Machura}
\affiliation{Institute of Physics, University of Augsburg,
Universit\"atsstrasse 1, D-86135 Augsburg, Germany}
\affiliation{Institute of Physics, University of Silesia,
P-40-007 Katowice, Poland}
\author{M. Kostur}
\affiliation{Institute of Physics, University of Augsburg,
Universit\"atsstrasse 1, D-86135 Augsburg, Germany}
\author{P. H\"anggi}
\affiliation{Institute of Physics, University of Augsburg,
Universit\"atsstrasse 1, D-86135 Augsburg, Germany}
\author{P. Talkner}
\affiliation{Institute of Physics, University of Augsburg,
Universit\"atsstrasse 1, D-86135 Augsburg, Germany}
\author{J. \L uczka}
\affiliation{Institute of Physics,  University of Silesia,
 P-40-007 Katowice, Poland}

\date{\today}
\begin{abstract}
A quantum Smoluchowski equation is put forward that consistently
describes thermal quantum states. In particular, it notably does
not induce a violation of the second law of thermodynamics. This
so modified kinetic equation is  applied to study {\it
analytically} directed quantum transport at strong friction in
arbitrarily shaped ratchet potentials that are driven by
nonthermal two-state noise. Depending on the mutual interplay of
quantum tunneling and quantum reflection these quantum corrections
can induce both, either a sizable enhancement or a suppression of
transport. Moreover, the threshold for current reversals becomes
markedly shifted due to such quantum fluctuations.
\end{abstract}

\maketitle

Brownian motors are small physical machines that operate far from
thermal equilibrium by  extracting energy  fluctuations to
generate work against external loads \cite{ratch}. They present
the physical analogue of biomolecular motors that direct
intracellular transport and control motion and sensation in cells
\cite{RMPJUL}. In contrast to these molecular bio-motors, however,
the molecular sized physical engines necessitate -- depending on
the nature of particles to be transported and their operating
temperature -- a  description that accounts as well for quantum
features such as tunneling and quantum reflection. For this class
of  quantum Brownian motors, recent theoretical studies
\cite{QBM,QBMTH} have predicted that the transport  becomes
distinctly modified as compared to its classical counterpart. In
particular, innate quantum effects such as tunneling induced
current reversals, power-law like quantum diffusion transport
laws, and quantum Brownian heat engines have been observed  with
recent, guiding experiments that  involve either arrays of
asymmetric quantum dots \cite{linke} or  cell-arrays composed of
different Josephson junctions \cite{majer}.

Because of the mutual interplay of quantum mechanics, dissipation
and non-equilibrium driving, the theoretical description of such
nonequilibrium, dissipative quantum Brownian motor devices  is
notoriously difficult. The present state of the art of the theory
is characterized by specific restrictions such as e.g. an
adiabatic driving regime, a tight binding description, a
semiclassical analysis, or combinations thereof \cite {QBM,QBMTH}.
As such, the study of quantum Brownian motors is far from being
complete and there exists  an urgent need of  further
developments. The analytic study of  quantum Brownian transport
for  {\em arbitrarily shaped} spatially periodic ratchet
potentials presents such a challenge. This goal is addressed here
within the strong friction regime, where the underlying quantum
dynamics can be modeled by a recently put forward, ingenious
quantum generalization of Smoluchowski dynamics \cite{anker,pech}.

Classically,  a system coupled to a thermal bath  at temperature
$T$ is described in terms of Langevin equations or corresponding
Fokker-Planck  equations \cite{HT}. For a Brownian particle this
yields the Kramers equation  which in  the strong friction limit
reduces to the Smoluchowski equation. In quantum statistical
physics the description of Brownian motion dynamics is distinctly
more intricate; it has been worked out, however, in some detail
within limited generality using e.g. the assumption of a linear
bath dynamics or  a weak coupling limit. For the latter case,
quantum master equations, e.g. of Lindblad form, have been derived
\cite{Ingold,ali}.

{\it Quantum Smoluchowski dynamics}.-- Recent work within the
strong friction  limit shows that  quantum Brownian motion can be
described by  a generalized Smoluchowski equation that accounts
for leading quantum corrections \cite{anker,pech}. For a particle
of mass $M$ moving in the potential $V(x)$, Ankerhold {\it et al.}
proposed   a quantum Smoluchowski equation (QSE) for the diagonal
part of the density operator $\rho(t)$, i.e.  the rate of change
of the
 probability density $P(x, t) = \langle x\vert \rho(t) \vert x\rangle$
in  position space $x$ assumes the form \cite{anker}:
\begin{equation} \label{S1}
\gamma M \frac{\partial}{\partial t} P(x,t) =
\frac{\partial }{\partial x} V'_\text{eff}(x) P(x,t)
+  \frac{\partial^2}{\partial x^2} D_\text{eff}(x) P(x, t),
\end{equation}
where $\gamma$ denotes friction. The effective potential reads
\begin{equation} \label{effpot}
V_\text{eff}(x) = V(x) + (1/2) \lambda V''(x),
\end{equation}
wherein the prime denotes the derivative with respect to the coordinate $x$.
The prominent parameter
\begin{equation} \label{lam}
\lambda = (\hbar / \pi M \gamma) \ln (\hbar \beta \gamma / 2 \pi),
\quad \beta = 1/k_B T,
\end{equation}
describes  quantum fluctuations in  position space and $k_B$ is
the Boltzmann constant. The effective diffusion coefficient reads
\cite{anker}
\begin{eqnarray} \label{D1}
D_\text{eff}(x)=D_\text{Ank}(x) = \beta^{-1} [1 + \lambda \beta V''(x)].
\end{eqnarray}
Note that eq. (\ref{S1}) is valid whenever   $k_BT \ll \hbar
\gamma$.

This so derived quantum-Smoluchowski equation exhibits, however, a
disturbing shortcoming: In clear contradiction to the validity of
the second law of thermodynamics, eq. (\ref{S1}) yields  for an
arbitrary, {\it asymmetric} periodic ratchet potential $V(x)$ of
period $L$ at {\it zero} external  bias a {\it non-zero},
stationary average velocity $\langle v \rangle  = J L$  (or
equivalently, a non-vanishing probability current $J$). This is
so, because the expression for the current reads
\begin{equation} \label{genJ}
\langle v \rangle = \frac{L}{\gamma M} \; \frac{ 1 - \exp [ \Psi
(L) ] }{\int_{0}^{L} \mathrm{d} x D_\text{Ank}^{-1}(x)\exp [ - \Psi
(x)] \int_{x}^{x + L} \mathrm{d} u \exp [ \Psi (u)]} \;,
\end{equation}
where $ \Psi(x)=\int_{0}^{x} \mathrm{d}u V'_\text{eff}(u) / D_\text{Ank}(u)$
 upon inspection is non-periodic with
 $\Psi(L)\neq 0$. Thus, a finite
stationary drift emerges; i.e. a Maxwell demon seemingly is at
work at stationary, thermal equilibrium.

{\it Demon-free quantum-Smoluchowski dynamics}.-- Next, we put
forward a clear-cut modification of the above quantum-Smoluchowski
equation which does not cause such a fake perpetual motion
phenomenon. First, we observe from theory \cite{anker} that the
{\it leading} strong friction quantum correction involves the
second order derivative of the potential $V(x)$, see (\ref{D1}).
Following  prior works \cite{anker,pech} we shall consistently
neglect (in the high friction limit) higher order contributions in
${\lambda}$, which in fact would involve also higher order
derivatives of the potential. This new, modified quantum
Smoluchowski equation (M-QSE) is derived from the following set of
construction criteria: We seek a new diffusion coefficient that
(i), in leading order reproduces the previous result in Refs.
\cite{anker,pech}, and (ii), does not exhibit a Maxwell demon
behavior, i.e. the modified dynamics yields in thermal equilibrium
a vanishing probability current, and additionally (iii), the
dynamics reproduces  the correct thermal quantum position
probability for strong friction \cite{gwt}. The construction
criteria (ii) of zero flux together with the correct leading order
result for the thermal position probability in (iii) then fixes
the form of the diffusion function $F[V''(x)] =a^{-1}[1 - b
V''(x)]^{-1}$ uniquely. The two constants $a$ and $b$  read
explicitly $a=\beta $ and $b=\lambda \beta$.

Upon an expansion of $F[V''(x)]$ into a series in $\lambda$ the
two diffusion functions  do  coincide  in first order with respect
to the  quantity $\epsilon (x) = \vert\lambda \beta V''(x)\vert <
1$, as required by the condition  in (i). Therefore,  this
improved modified quantum-Smoluchowski equation (\ref{S1}) is
given by a modified diffusion, reading
\begin{eqnarray}
\label{DD2}
D_\text{eff}(x)=D_\text{mod}(x) =\beta^{-1}[1 - \lambda \beta V''(x)]^{-1}\;.
\end{eqnarray}
Note that from a mathematical viewpoint our thermal M-QSE-dynamics
assumes the form of  a Pad\'e-like, non-perturbative result in
place of (\ref{D1}). The  thermal quantum-Smoluchowski stochastic
dynamics in this strong friction limit is thus equivalent to
classical  Brownian dynamics within the effective potential
(\ref{effpot}) and the new, state-dependent diffusion coefficient
given in (\ref{DD2}). The
 corresponding (M-QSE) Langevin equation reads in the Ito-representation \cite{HT}
\begin{equation} \label{Lan1}
\gamma M \dot{x} = -V'_\text{eff}(x) + \sqrt{2\gamma M D_\text{mod}(x)} \; \xi(t)
\;,
\end{equation}
where the dot denotes the time derivative and
$\xi(t)$ is (classical) Gaussian white noise of vanishing mean and
 correlation $<\xi(t)\xi(s)> =\delta (t-s)$.
The above  scheme  is close in spirit with the approximation
method of colored noise driven dynamics in terms of corresponding
effective Markovian processes \cite{colored}.


{\it Quantum Brownian motor transport}.-- A finite  transport emerges
when the system operates far from thermal equilibrium \cite{ratch}.
In the present context, we investigate overdamped, quantum Brownian motors
\cite{QBM,QBMTH} with the quantum  fluctuations characterized by
the parameter $\lambda$ in  (\ref{lam}). To this aim, we complement
the thermal quantum dynamics in eq. (\ref{Lan1}) with a slowly waggling
nonthermal, deterministic or random force $\eta(t)$, i.e.
\begin{equation} \label{Lan}
\gamma M \dot{x} = -V'_\text{eff}(x) + \sqrt{2\gamma M D_\text{mod}(x)}
+ \eta(t) \;.
\end{equation}
In dimensionless form we then obtain
\begin{eqnarray}  \label{aa}
 \dot y = -W'_\text{eff}(y) + \sqrt{2 \mathcal{D}_\text{mod}(y)} \; \hat \xi (s)
+ \hat\eta(s),
\end{eqnarray}
where the position of the Brownian motor  is scaled as $y= x/L$,
time is re-scaled as $s=t/\tau_0$, with the characteristic
time scale reading $\tau_0 = M \gamma L^2/\Delta  V$
(the barrier height $\Delta V$ is the difference between the maximal and minimal values of $V(x)$).
During this time span, a classical, overdamped particle moves a distance
of length $L$ under the influence of the constant force $\Delta V/L$.
The  effective  potential is
$W_\text{eff}(y) = W(y) + (1/2) \lambda_0 W''(y)$, where the re-scaled potential
$W(y) =  V(x)/ \Delta V = W(y+1)$
possesses a unit period  and a unit barrier height. The dimensionless parameter
$\lambda_0 = \lambda /L^2$ describes  quantum fluctuations
over the characteristic length L. For example, the value $\lambda _0 = 0.01$ means that,
roughly speaking,  the
difference between quantum and classical fluctuations of the position of
the Brownian particle  is significant over distances of the order $\sqrt{\lambda_0} L = 0.1 L$.
The re-scaled diffusion function $\mathcal{D}_\text{mod} (y)$ reads,
\begin{eqnarray} \label{DD3}
\mathcal{D}_\text{mod}(y) =\beta_0^{-1}[1 - \lambda_0 \beta_0  W''(y)]^{-1}\;.
\end{eqnarray}

The dimensionless,
inverse temperature  $\beta_0 = \Delta V/k_B T$
is a  ratio of the activation energy in the non-scaled
potential and the thermal energy. The re-scaled
Gaussian white noise is $\hat\xi(s)= ( L/\Delta V) \xi(t)$ and
 the re-scaled, nonthermal
force reads  $\hat\eta(s) = (L/\Delta V) \eta(t)$.

As a specific realization, we next consider nonthermal
fluctuations modeled by  Markovian, two-state noise, $\hat\eta
(s)=\{-a, a\}$, that switches with a rate
 $\nu$ between the levels $a$ and $-a$.
This problem can be  solved  analytically
in the adiabatic limit, i.e. if
$\nu \to 0$. In this limit  the stationary averaged dimensionless velocity reads
$\langle  \dot y \rangle = J  = (1/2)   [J(a)+J(-a)]$, where
\begin{equation}
J(a) =  \frac{1-\exp (- \beta_0 a)}
{\int\limits_0^1
 dy \;
\mathcal{D}_\text{mod}^{-1}(y) \exp [-\beta_0\Psi(y,a)]
\int\limits_y^{y+1} dz \; \exp [ \beta_0\Psi(z,a)]}
\end{equation}
and
\begin{eqnarray}
\Psi(y,a) &=& W(y) + (1/2) \lambda_0 W''(y)
-(1/2) \lambda_0 \beta_0 [W'(y)]^2 \nonumber\\
&-& (1/4)\lambda^2_0 \beta_0 [W''(y)]^2 + a \lambda_0 \beta_0
 W'(y) -ay.
\end{eqnarray}
Its classical behavior, i.e. $\lambda_0 = 0$, has been studied  in
Refs. \cite{kul}.
\begin{figure}[htbp]
\centerline{\includegraphics[width=8cm]{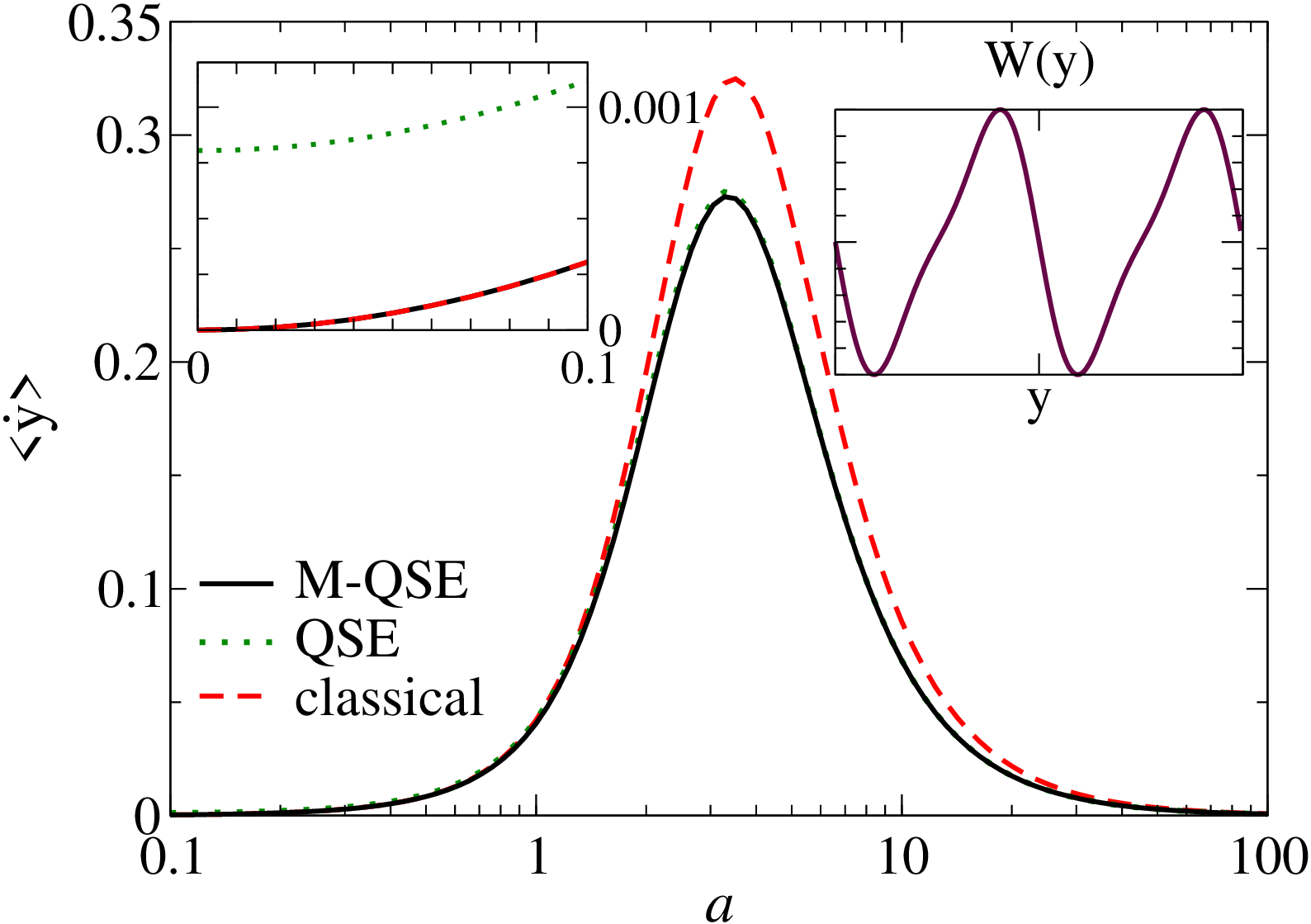}}
\caption{\label{fig1} Stationary velocity $\langle \dot{y}
\rangle$ {\it vs.} the two-state noise amplitude $a$  for both, a
strongly damped quantum Brownian motor (solid) and its classical
counterpart (dashed).  The  ratchet potential of unit barrier
height, $W(y) = - W_0 [\sin (2 \pi y) + 0.25 \sin (4 \pi y)]$,
with $W_0=0.454...$, is depicted with the inset. The theory
(dotted) in Ref. \cite{anker} yields a nonphysical (although
small) quantum-Maxwell demon behavior at small noise amplitudes
$a$; further away from equilibrium ($a>0.5$) the QSE and the M-QSE
predictions practically coincide within line-thickness. The chosen
dimensionless inverse temperature is $\beta_0 = 5$. }
\end{figure}

The influence of the quantum corrections is presented with Figs.\
1--3.   The role of quantum noise enters via two functions: The
effective potential $W_\text{eff}(y)$ and the effective diffusion
function $\mathcal{D}_\text{mod}(y)$. The quantum correction to the
potential depends logarithmically weakly on temperature. The
crucial correction  stems from the diffusion  which increases as
the temperature decreases. The prominent quantum effects appear
for lower temperatures. In Figs.\ 1--3, we take for the re-scaled
quantum fluctuations a parameter value of $\lambda_0 = 10^{-4}
\ln(10^3 \beta_0)$. This choice assures that the quantum
Smoluchowski regime is fully valid down  to low temperatures of
the order $\beta_0 \approx 10$. In Fig. 1 we depict the current
{\it vs.} the dichotomic noise level  $a$. We deduce that the
quantum corrections reduce  the absolute value of the current (the
maximal absolute quantum correction is $\vert \lambda_0 \beta_0
W''(y)\vert = 0.202$). Note that the current value approaches zero
for a vanishing noise amplitude $a \rightarrow 0$ (solid line)
and, as well, for very large amplitude $ a \rightarrow \infty$.
The modification of the diffusion coefficient turns out to be
essential for small amplitudes $a$ of the non-equilibrium
two-state noise; this regime describes the near-equilibrium
behavior with the directed current approaching zero. In clear
contrast, the use of the conventional quantum Smoluchowski
equation (QSE) in (\ref{D1}) (dotted) yields a nonphysical,
(although small) positive  current value. It should be pointed
out, however, that far away from equilibrium (for $a>0.5$) the two
forms of quantum-Smoluchowski dynamics yield practically identical
results.

The  analytic  expression for the current allows one to study {\it
arbitrarily shaped} ratchet profiles. As an example we consider
the more complex shaped asymmetric periodic potential
\begin{eqnarray}  \label{W(x)}
        W(y)= W_0 \{\sin(2\pi y)+0.4 \sin[4\pi(y-0.45)] \nonumber\\
     + 0.3 \sin [6\pi (y-0.45)]\},
\end{eqnarray}
where $W_0= 0.371$ normalizes the barrier height to unity, see
inset in Fig.\ 2.
\begin{figure}[htbp]
\centerline{        \includegraphics[width=8cm]{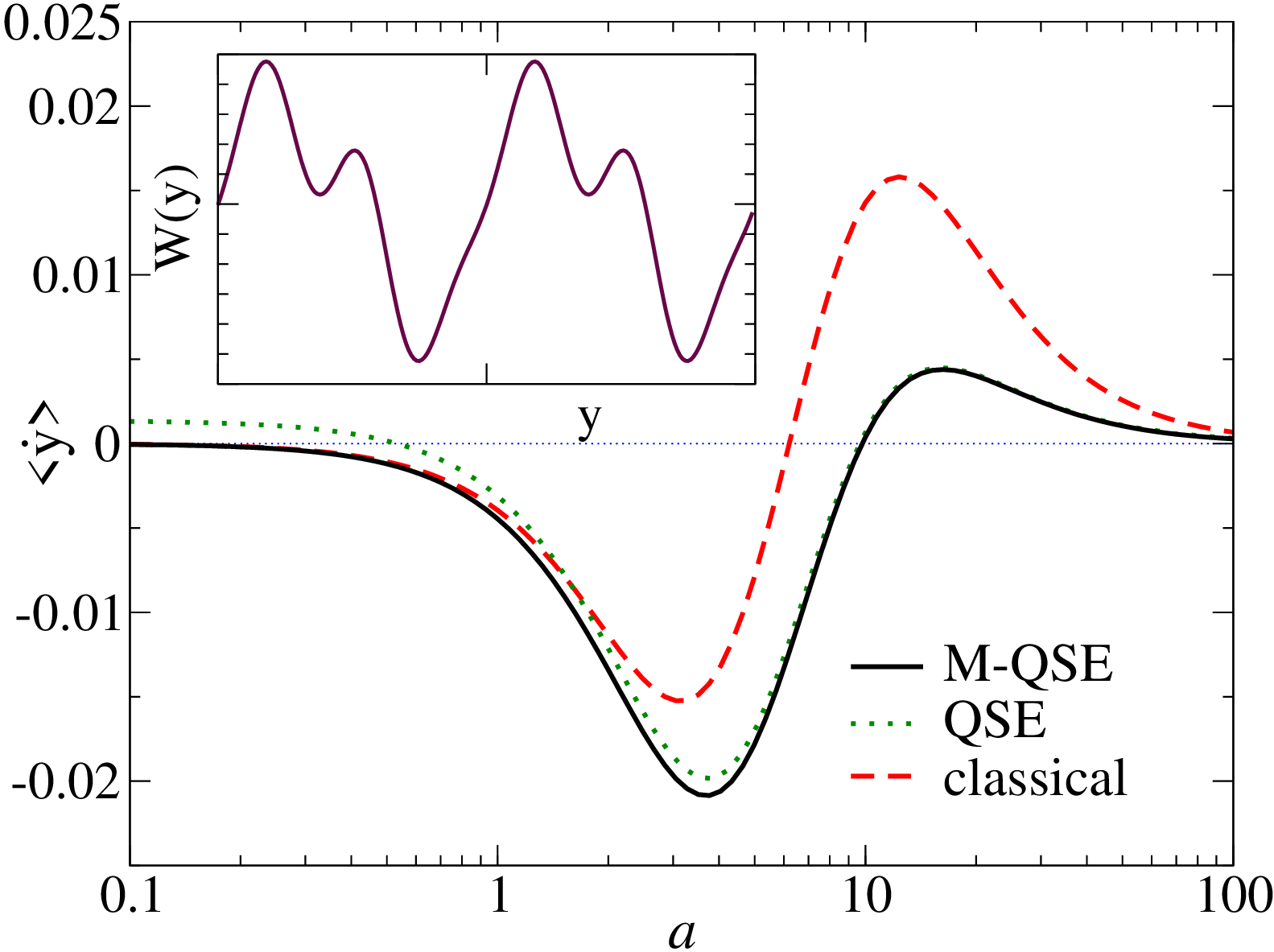}}
\caption{\label{fig2} The dependence of the stationary current
$\langle \dot{y} \rangle$ {\it vs.} the two-state noise level $a$
is depicted for the potential (\ref{W(x)}) (inset),  for both, the
modified quantum-Smoluchowski (M-QSE) theory  (solid), the
conventional quantum-Smoluchowski (QSE) theory (dotted) and the
classical case (dashed), respectively, for an inverse
dimensionless temperature  $\beta_0 = 2$. Note that the maximal
absolute correction is actually rather small: $\vert \lambda_0
\beta_0 W''(y)\vert = 0.06$. }
\end{figure}
This ratchet potential exhibits an intriguing  current reversal
{\it vs.} the noise amplitude $a$. The maximal absolute quantum
correction is $\vert \lambda_0 \beta_0 W''(y)\vert = 0.06$.  There
occur two regimes: one regime of small noise levels {$a$} for
which the amplitude of the quantum current is enhanced and one at
larger noise amplitudes where the classical current exceeds its
quantum counterpart. A most salient intermediate regime occurs for
which the {\it classical current is positive while the quantum
current remains
  negative}. The point of the physically relevant quantum current
reversal is shifted towards larger noise levels. Use of the
quantum-Smoluchowski diffusion theory of Refs. \cite{anker,pech}
yields a fake, positive-valued current at small dichotomic noise
strength (dotted line), being accompanied by a non-physical (!)
current reversal, see in Fig.\ 2. We find again that the expected
convergence between the two theories occurs far away from thermal
equilibrium.

\begin{figure}[htbp]
\centerline{        \includegraphics[width=8cm]{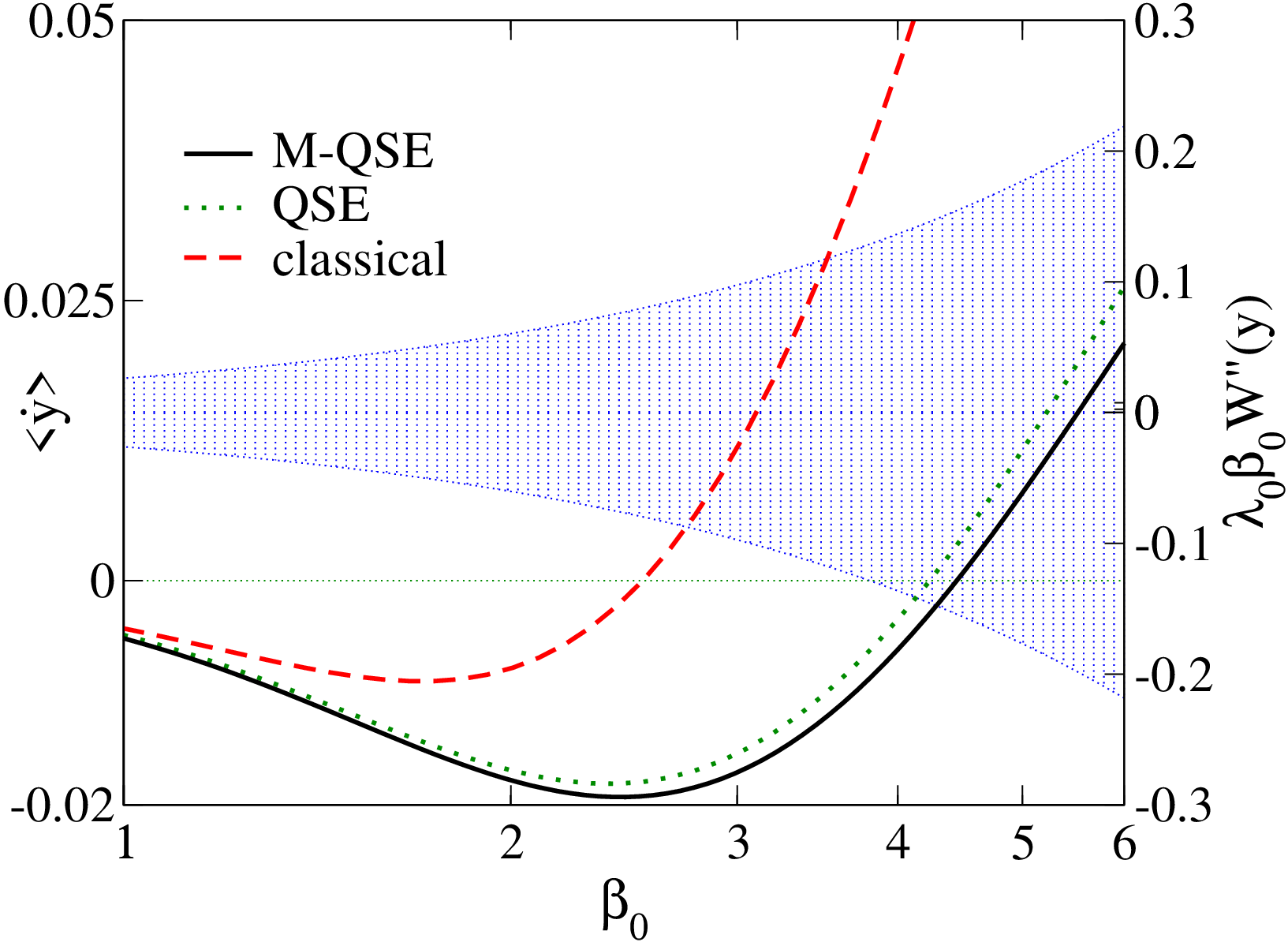}}
\caption{\label{fig3} The directed quantum noise-induced transport
$\langle \dot{y} \rangle$  of the quantum Brownian motor (solid)
{\it vs.} the dimensionless inverse temperature $\beta_0$ for the
ratchet potential (\ref{W(x)}), see inset in Fig.2, is compared
with its classical limit (dashed) and the conventional quantum
theory in (\ref{D1}) (dotted). The dichotomic noise level is set
at $a = 5$. The variations of the  quantum corrections are
depicted with the dotted background. }
\end{figure}
In Fig. 3   we elucidate the role of temperature. Equation
(\ref{DD3}) shows that the quantum corrections increase
monotonically as temperature decreases. We must refrain ourselves,
however, from analyzing the limit of extreme  low temperature.
This is so, because the quantum corrections then grow too large,
causing the diffusion  to pass from  positive to nonphysical,
negative values upon exceeding the threshold value $1$; clearly,
the strong friction quantum theory is valid only  below this
threshold. In fact, for correction values close to threshold the
non-diagonal, density matrix elements assume nonzero decoherence
values that can no longer be neglected with the
quantum-Smoluchowski theory. Fig. 3  depicts  these increasing
quantum corrections  with decreasing temperature. At high
temperatures the interplay between reflection and tunneling
transmission causes a larger (in absolute value) quantum current;
upon crossing the point of classical  current reversal this
behavior is interchanged. At even lower  temperatures quantum
corrections cause a smaller current value.

{\it Conclusions}.-- By use of a distinct modification of
quantum-Smo\-lu\-chow\-ski theory we have developed  a strong
friction quantum approximation that is in agreement with both
thermal equilibrium statistics and --  above all -- with the
second law of thermodynamics. This so obtained, modified quantum
theory can be applied for far from equilibrium  transport where it
facilitates closed form expressions (in terms of quadratures) for
directed, quantum Brownian motor transport. Our  tractable results
hold true away from the semiclassical limit and, additionally, can
readily be applied to experimentally, arbitrarily shaped ratchet
profiles. Note that this presents an important advance over prior
studies of quantum ratchets \cite{QBM,QBMTH} that often require
the use of manageable,  stylized potential forms. Our
investigation additionally manifests a rich spectrum of quantum
Brownian motor behaviors, exhibiting both, quantum induced
enhancement and suppression of transport, as well as shifted
current reversals.

These novel features can advantageously be put to work for quantum
ratchets on the micro- and nanoscale \cite{ratch}. Moreover, the
structure of our quantum-Smoluchowski dynamics can be generalized
to higher dimensional overdamped situations as e.g. for quantum
noise-induced directed transport on surfaces. In particular, our
method and these quantum ratchet signatures  can be utilized to
optimize transport properties in superconductors by controlling
the motion of vortices and magnetic flux quanta
\cite{nori,science}.

Financial support by the  DAAD (L.M.), the Foundation for Polish
Science (P.H.), the DFG via grant HA-1517/13-4 and the ESF
(Programme Stochdyn) is gratefully acknowledged.

\end{document}